\documentclass[12pt]{article}
\usepackage{graphicx}
\renewcommand{\baselinestretch}{1.2}

\begin{document}

\title{A new nestedness estimator in community networks}
\author{Gilberto Corso$^1$, Aderaldo I. L. de Araujo$^2$\\
 Adriana M. de Almeida$^3$  \\
1. Departamento de Biof\'isica e Farmacologia, Centro de Bioci\^encias, \\
Universidade Federal do Rio Grande do Norte \\
UFRN - Campus Universit\'ario, Lagoa Nova,\\
CEP 59078 972, Natal, RN, Brazil  \\
Phone 84 3215 3419 Fax 3211 9204 \\
email: corso@cb.ufrn.br \\
2. Departamento de F\'isica Te\'orica e Experimental, \\
Universidade Federal do Rio Grande do Norte \\
UFRN - Campus Universit\'ario, Lagoa Nova,\\
CEP 59078 972, Natal, RN, Brazil  \\
3. Departamento de Bot\^anica, Ecologia e Zoologia, \\
Centro de Bioci\^encias, Universidade Federal do Rio Grande do Norte \\
UFRN - Campus Universit\'ario, Lagoa Nova, \\
CEP 59078 972, Natal, RN, Brazil  }

\maketitle

\newpage

\begin{abstract}
A recent problem in community ecology lies in defining structures
behind matrices of species interactions.  
The interest in this area is to quantify the nestedness degree of the matrix
after its maximal packing.   In this work we evaluate nestedness
 using the sum of all distances of the occupied sites to the vertex of the matrix.
 We calculate the distance for two artificial matrices with the same size and occupancy:
a random matrix and a perfect nested one. Using these two benchmarks we develop
a nestedness estimator.  The estimator is applied to a set of $23$ real
networks of insect-plant interactions.

\end{abstract}

\hspace{-0.6cm}{\bf Keywords:} nestedness, networks, insect-plant interactions

\section{Introduction}

Networks have been widely used to describe systems in a multitude of disciplines, such as genetic networks, protein networks or the Internet. In ecology, networks are mainly used to visualize and describe food webs. But not only trophic interactions are the focus of attention. In the last years researchers show a growing interest in the study of other species interactions such as parasitism (Vázquez et al., 2005), scavenger species (Selva and Fortuna, 2007) and mostly mutualism (Bascompte and Jordano, 2007 and references therein). Studies on mutualistic food webs focus on specific pairwise interactions between a plant and an animal and how they are shaped by a community context, either in a single locality, or geographically (Bascompte and Jordano, 2007).

Pairwise interactions can be described in the form of a bipartite graph or an interaction matrix. These webs are characterized by nodes that represent species or species groups and observed interactions are drawn as links that, when not binary, can render their intensity or frequency in graded thicknesses. In the interaction matrix, links are represented as nonzero cells on the intersection of a row and column. According to Almeida-Neto et al.,2007, bipartite webs do in fact offer several advantages of their own: first, they are often fully resolved, without the problems of uneven resolution which haunt the analysis of complete webs. Second, all links are of a single kind of ecological interaction (e.g. mutualism), which ensures structural integrity as well as similar ecological and evolutionary processes throughout the entire assemblage.

The most studied structure within a bipartite graph is the nested pattern of species interactions, although other structures are also possible (Prado et al. 2006; Almeida-Neto et al., 2007). In nested assemblages, plants with few interactions are related only with generalist animals; conversely, specialized animals are found related to plants with many links, that is, with large associated faunas. Moreover, generalists in one species set tend to interact with generalists in the other, forming a dense core of interactions (Prado et al., 2006; Bascompte and Jordano, 2007).

A nested structure is very cohesive and stable. The fact that few species are involved in many interactions (functional redundancy), poses the community with the possibility for alternative routes if some interactions disappear (Bascompte and Jordano, 2007). A nested structure is also quite robust: it is less prone to sampling bias than number of species and links (Nielsen and Bascompte, 2007) and not  generated by the random combination of sets of plants and animals solely in proportion to their different abundances as previously thought (Prado et al., 2006).

Recently, a large series of mutualistic interaction assemblages have shown a significantly nested structure (Bascompte et al. 2003). Substantial effort has been done in developing various measures and forms of calculating nestedness (Atmar and Patterson, 1993; Guimaraes and Guimaraes, 2006; Rodríguez-Gironés and Santamaría, 2006; Almeida-Neto et al., submitted).
The most commonly used nestedness metric is the “Nestedness Temperature Calculator”, or “Nestcalc”, used to calculate nestedness in binary matrices (Atmar and Patterson, 1993). The nestedness from this algorithm has a problem, though: the absolute value of the nestedness temperature is dependent on matrix size and fill. Some studies show that the nestedness temperature of randomly assembled matrices increases with network size and attains its maximum value for intermediate fills (Rodríguez-Gironés and Santamaría, 2006; Almeida-Neto et al., submitted). So, smaller networks need lower temperature than larger ones do to be significantly nested (Nielsen and Bascompte, 2007).

From a mathematical point of view, the central object in the discussion about nestedness is a matrix of zeros and ones. The ecologist in the field interpret this matrix as a table where she (he) marks a cross at the $i$ column and $j$ row each time a species of group one $i$ (e.g. plant) is related to group two $j$(e.g. insect). In order to visualize nestedness in the studied ecological community the ecologist has to rank rows and columns of the table. In fact, each time one row (or column) is permuted to another row (or to column), the interactions among species of groups one and two do not change. For a matrix of $L_1$
species in group one and $L_2$ species in group 2, there are
$L_1 \! \times L_2 \!$ possibilities to represent the matrix following different permutations of rows and columns. Each one of these possibilities is just a different visualization of the same network structure.

Ranking rows and columns is a very practical option to visualize nestedness in a interaction matrix. When we rank the elements of a matrix we choose one of the $L_1 \! \times L_2 \!$ possibilites, that one where the elements of the matrix are the most packed. In other words, we choose the representations where the elements of the matrix are as closest as possible from the $i=1$, $j=1$ corner. In the literature . Packing procedure is a previous step before the evaluation of a nestedness index (Atmar and Patterson, 1993). In this article we also pack the matrix before the evaluation of our nestedness index.

We introduce here a new nestedness measure applied to digraphs
originated from ecological data. However, the method is more general than its predecessor and
can be naturally applied to graphs (networks) other than digraphs.  In section $2$ we describe the
formal objects used in this article: adjacency matrix, Manhattan distance
in a matrix, projection of a generic matrix into the unit square lattice, packing
process, random matrix, maximum nested matrix  and
nestedness estimator. In section $3$ we apply the nestedness estimator to a set
of insect-plant herbivory networks extracted from the community ecology literature. In section
$4$ we summarize the article, point out potential applications of the method and
give the final words.

\section{A measure of nestedness}

In this section we introduce the concept of distance in a matrix to characterize the
nestedness of digraphs. In order to fix
the notation we call digraph an object $D$ formed by two sets of vertices
 $D_1$ and $D_2$ and a set of links between these two sets. The digraph is
completely described by the adjacency matrix, $M$, of size $L_1 \times L_2$, where $L_1$ and
$L_2$ are the number of elements of $D_1$ and $D_2$, respectively. By definition $M_{i,j}=1$ if
there is a link between vertices $i$ of $D_1$ and $j$ of $D_2$
  and $M_{i,j}=0$ if $i$ and $j$ are
not linked. It is useful to visualize $M$ as a $L_1$ versus $L_2$ lattice
with empty (zero) or full (one) sites.
 Moreover, the number of links of a vertex $l$ is $k_l$ and the
distribution of links of $D_1$ and $D_2$ is $P_{l_1}$ and $P_{l_2}$ respectively.

In Ecology, the field data corresponding to the digraph is composed by
two  sets of species and the corresponding links (interactions) between them. As we pointed out in the introduction, the  standard
procedure in this area  consists in packing the adjacency matrix of the
data. The packing is performed in the following way: the link distributions
$P_{l_1}$ and $P_{l_2}$ are ordered such that the most connected species go to the
first position of the matrix. In this way the matrix $M$ shows more ones close to $i=1$ and
$j=1$ corner and zeros at the opposite corner, $i=L_1$ and $j=L_2$. From the matrix
point of view, the packing process consists in  replacing lines and columns until
$P_{l_1}$ and $P_{l_2}$ are ordered. We emphasize that since the packing process
 do not change the links between species it does not alter the phenomenology
underlying the network.

The idea behind packing the matrix $M$ is
to better visualize network nestedness. In addition, nestedness is related with
the dispersion of ones and zeros after the packing process. A very nested matrix is
one that, after packing, has a minimal mixing of ones and zeros. Using
a lattice analogy, a very nested lattice  shows a minimum of holes.

 To introduce distance properties in the original matrix $M$ we map it
into a Cartesian space. In order to avoid distortions we map
the $L_1 \times L_2$ matrix to the unit square.
To perform this task the cell elements $(i,j)$ assume
 the positions  $x_i$ and $y_i$ that are done by:
$$
        x_i = (i-1)/L_1 + 1/(2 L_1)
$$
\begin{equation}
              y_j = (j-1)/L_2 + 1/(2 L_2)
\label{xey}
\end{equation}

In this article we use the Manhattan distance because it is broadly  employed
to measure matrix distances. Euclidean distance is
used for estimating distance between elements apart in continuum space, which
is not the case here. In fact, in the context of abstract metric spaces
(Courant, R. and Hilbert, D., 1937)
set of distances $d_{\chi} = (x^{\chi}+y^{\chi})^{1/\chi}$ that depends on
the parameter $\chi$, the case $\chi=1$ corresponds to the Manhattan distance
and the case $\chi=2$ to the Euclidean distance.

 We define the occupancy number $\rho$ as the fraction of occupied sites in the adjacency matrix. For $N$ the total number of ones in $M$ we have $\rho=\frac{N}{L_1 L_2}$.
To quantify nestedness of a given matrix $M$ we use two matrix benchmarks with the
same $L_1$, $L_2$ and $\rho$: the maximal nested matrix $\tilde M$ and the random
matrix $M_{rand}$.
The maximal nested matrix is constructed in such a way that it has
no holes and its elements are as close as possible to the $(1,1)$ corner. We
construct $\tilde M$ filling the elements along equidistant diagonals to
$(1,1)$. In fact all elements along the same diagonal have
the same distance to the $(1,1)$ corner.  The
construction of $\tilde M$ is the following: the first element occupied is $(1,1)$,
after that it comes $(1,2)$ and $(2,1)$, followed by $(1,3)$, $(2,2)$ and $(3,1)$, etc.
 Figure \ref{fig1} illustrates the optimal filling strategy to build $\tilde M$.  In contrast, the random matrix is constructed in such a way that all its elements are
uniformly occupied with the same probability $ p = \rho$.
 The maximum packed matrix of the species interaction will be in-between these two.

We use the Manhattan distance to evaluate the distances of the filled elements $x_i,y_j$
(the distances are defined for the matrix elements projected into the unit square in the
cartesian plane). We call $d$ the sum over the distances of all the elements
of the matrix projected into the unit square, that means, $d=\sum d_{i,j}$ for
$d_{i,j}=x_i+y_i$. In order to define the nestedness estimator we introduce
two additional distances: the total
distance of the artificial matrices $\tilde M$ and $M_{rand}$.
We note that $\tilde M$ has the smallest total distance among all the
lattices with the same $\rho$ and we call $d_{min}$ its total distance, while
the total distance of $M_{rand}$ is $d_{rand}$. Consider a sample of $N$ points ramdomly
distributed along the unit square, the expected value of the distance to the origin, $\mu$,
 is the Manhatann distance from the origin to the center of the
square of size $1$, that means, $\mu = 1$.  Therefore:
\begin{equation}
 d_{rand} = \sum_{k=1}^{N} d_{i,j} = N \mu = N
\label{eqrand}
\end{equation}

 To get an insight about distances in $M$ we  start exploring the
behavior of $d_{min}$ and $d_{rand}$ against occupancy $\rho$ in
 figure \ref{fig2}. We use in this picture $L_1=L_2=20$.
 As expected, the distances follow the relation
$d_{min} \leq d_{rand}$. The
total distance $d$ for any matrix, after the packing process,
shows the property:
\begin{equation}
 d_{min}  < d  <  d_{rand}.
\label{eq333}
\end{equation}
In fact, $d_{min} <d $ since $d_{min}$ is derived from an artificial matrix whose components,
by construction, have the minimal distance to the origin. Otherwise, $d< d_{rand}$
because $d$ is derived from a packed matrix, and in the packing process the matrix
reduces the distances of their elements when compared with a similar random matrix.

The distance as defined above depends on the matrix size and the
occupation. In fact, the total distance observes the relation $d \propto L^2$ for
a given $\rho$ and  the relation $d \propto \> N^2$ for a constant $L$.
This behavior can be visualized in figures \ref{fig2} and \ref{fig3}. In order to have a $N$ free
nestedness index of the system we define the nestedness index $\eta$ as follows:
\begin{equation}
               \eta = \frac{d - d_{min}}{d_{rand} -d_{min}}
\label{eqd}
\end{equation}
  We emphasize that $d_{min}$ and $d_{rand}$
are computed  over a artificial matrix with the same $L_1$, $L_2$ and $\rho$ of
the original system.  In the next section we test $\eta$ over a set of digraphs
from  the context of community ecology and discuss the results.

\section{Results}

In this chapter we select a set of $23$ insect-plant herbivory networks in the literature and apply the
nestedness index we develop in this article.  In table \ref{tab1} we enumerate the set of networks
with its main properties:  the occupancy $\rho$, size $L_1$ and $L_2$, the nestedness estimator $\eta$,
the temperature $T$ (according to Atmar and Patterson, 1993) and the reference of the network.
 A visual inspection of the table does not reveal correlation between  $\eta$ and $T$. In fact,
a linear correlation analysis between the two variables revels no significant correlation
($R = 0.19$ and $p=0.18$).

The range of values of our estimator is $0.17 < \eta < 0.83$ and the
average value is $\bar \eta = 0.45$. In contrast, the usual
temperature estimator have the range $6.8 < T < 43$ and average
value $\bar \rho = 18.6$.   In order to improve the visual intuition
about the problem we plot in figure $4$ four lattices of
insect-plant networks. It is clear in the figure that (a) and (b)
are highly nested, and that on the contrary, (c) and (d) are not
nested at all. This intuitive idea is corroborated by the $\eta$
estimator, but not by the temperature, the estimator of the two
initial matrices are 0.18 and 0.23 (low values), and of the last two
0.79 and 0.76 (high values). The temperature estimator, on the other
hand, shows an intermediary value in case (a), where our estimator
shows a low value and a very high temperature in (c) where our
estimator points for a very large value. In fact, figures (c) and
(d) have a large number of specialists, and in consequence the
matrix cannot be well nested. Our estimator corroborates this
observation. As a final remark concerning this set of figures, we
point that
 all the matrices in this figure are well packed,
that means, the number of elements of lines and columns are ordered. On the other hand, the nestedness calculator from Atmar and Patterson, 1993, usually fails in packing well the matrices.

From the observation of figure \ref{fig5} we see that the variables are correlated, in fact, a linear correlation analysis reveals $R=-0.74$, $t=5.0$ and $p=5.9 \times 10^{-6}$. An exponential correlation regression results $R=-0.86$, $t=6.3$ and $p=2.9 \times 10^{-6}$. Therefore the adjust of the data to the exponential curve is slightly better than the linear one. The dependence of a nestedness index to occupancy was already pointed in the literature (Gironés and Santamaría, 2006). At first, a relation between $\rho$ and $\eta$ seems intuitive: once the number of sites increase, in the average, they will be more nested after the packing process. We let for a future work a more carefull analysis of this point.

\section{Final Remarks}
In this work we develop a new nestedness estimator $\eta$ based on distances over the adjacency matrix of
the network. We think that this estimator will be useful in the methodological discussion involving
nestedness in community ecology. To make the method clearer to the reader we summarize
the algorithm to find $\eta$ in the following sequence of steps:

\begin{enumerate}

\item Evaluate the link distributions $P_{L_1}$ and $P_{L_2}$ of the adjascency matrix of
the network.

\item Pack the matrix, that means, permute lines and columns of the matrix in
order that $P_{L_1}$ and $P_{L_2}$ are ranked. This step defines a corner of nestedness.

\item Project the matrix into the unit square in order to avoid distortion due to the
diferences between the sizes $L_1$ and $L_2$ of the matrix.

\item Find the manhatann distance $d_{x,y}$ of all elements $x_i$ $y_i$ of the matrix and sum
to find the total distance of the elements of the matrix $d$.

\item Determine analitically the distance of the associated random matrix
with the same occupancy $\rho$:  $d_{rand}= N$, for $N$ the total number of
occupied elements of the matrix.

\item Determine computationally the distance of the asociated maximally
nested matrix with the same $\rho$:  $d_{min}$.

\item Finally, calcule the estimator $\eta = \frac{d - d_{min}}{d_{rand} -d_{min}}$.

\end{enumerate}

As estimated above, $0< \eta < 1$. In the limit $\eta \rightarrow 0$ the network is
completely nested and  $\eta \rightarrow 0$ corresponds to
the random limit. We tested our estimator for a set of $23$ insect-plant networks and
the data is summarized in table \ref{tab1}. An interesting result of our estimator is
that it depends on the occupancy number. This result is in agreement with the intuitive idea that
the matrix nestedness increases with its occupancy density.

The parameter $T$ is a very used nestedness
estimator in community ecology. This parameter, despite of its
popularity, is not well defined and present several problems (Fischer and Lindenmayer, 2002; Rodríguez-Gironés and Santamaría, 2006; Almeida-Neto et al., submitted).
 We are perfectly aware that our estimator will be compared with the usual temperature
estimator developed by Atmar and Patterson (1993). What we should do is
to show the good points of our method and let the methodological discussion
to  the scientific community. In this way we stress the strong
points of our method in the following:

\begin{enumerate}

\item Our algorithm is based on plain geometry and metric statements. In this way it is simple
and can be calculated with help of a easy computer program.

\item We have two benchmarks clearly defined: the total
distance of the random matrix and of the completely nested one.

\item We do not use any ad hoc parameter in the
equation that defines the nestedness estimator.

\item Our estimator gives a number between zero and one.

\item The visual inspection criterion of nestedness of empirical matrices
 agrees with our estimator.

\end{enumerate}

This paper opens a new perspective in the study of nestedness. We develop an
original index to measure nestedness that is based on direct metric analysis
of the matrix. Instead of considering the dispersion of elements around an artificial
isocline, we estimate directly the distances of all the matrix elements from
the packing corner. The nestedness of a matrix is a measure of how much the elements of
the matrix are close to the corner where the matrix is packed. We hope this paper to be usefull to improve the understanding of nestedness in the community ecology context.

\break \vspace{1cm} \textbf{\Large Acknowledgements} \vspace{0.5cm}

The authors thank M. Almeida-Neto for helpful comments on the
manuscript. Umberto Kubota, Graciela Valadares and Thomas Lewinsohn,
who made unpublished data available. The authors gratefully
acknowledge the financial support of Fapesp and CNPq, Brazil.
\vspace{1cm}

\textbf{\Large Bibliography}
\begin{description}

{\renewcommand{\baselinestretch}{2}\normalsize

\item Almeida-Neto, M. et al. 2007. On nestedness analyses: rethinking matrix temperature and anti-nestedness. Oikos 116: 716-722.

\item Atmar, W. and Patterson. B. 1993. The measure of order and disorder in the distribution of species in fragmented habitat. Oecologia. 96:373-382.

\item Bascompte, J. et al. 2003. The nested assembly of plant-animal mutualistic networks. PNAS 100: 9383-9387.

\item Bascompte, J. and Jordano, P. 2007. Plant-Animal Mutualistic Networks: The Architecture of Biodiversity. ARES 38:
567-593.

\item Buruga, J. H. and Olembo, R. J. 1971. Plant food preferences of some sympatric drosophilids of Tropical Africa. Biotropica 3: 151-158.

\item Claridge, M. F. and Wilson, M. R. 1981. Host plant associations, diversity and species-area relationships of mesophyll-feeding leafhoppers of trees and shrubs in Britain. Ecol. Ent. 6: 217-238.

\item Courant, R. and Hilbert, D. 1937. Methods of Mathematical Physics. Berlin: Julius Springer.

\item Dawah, H. A. et al. 1995. Structure of the parasitoid communities of grass-feeding chalcid wasps. J. An. Ecol. 64: 708-720.

\item Flowers, R. W. and Janzen, D. 1997. Feeding records of Costa Rican leaf beetles (Coleoptera: Chrysomelidae).
Florida Entomologist 80: 334-366.

\item Fischer, J. and Lindermayer, D. B. 2002. Treating the nestedness temperature calculator as a "black box" can lead to false conclusions. 99:
193-199.

\item Futuyma, D. J. and Gould, F. 1979. Associations of plants and insects in a deciduous forest. Ecol. Monog. 33-50.

\item Guimar\~aes, P. R. Jr. and Guimar\~aes, P.R. 2006. Improving the analyses of nestedness for large sets of matrices. Environmental Modelling and Software 21: 1512-1513.

\item Jermy, T. and Szentesi, A. 2003. Evolutionary aspects of host plant specialisation - a study on bruchids (Coleoptera: Brichidae). Oikos 101: 196-204.

\item Joern, A. 1979. Feeding patterns in grasshoppers (Orthoptera: Acrididae): Factors influencing diet specialization. Oecologia 38: 325-347.

\item Lewinsohn, T. M. et al. 2006. Structure in plant-animal interaction assemblages. Oikos 113: 174-184.

\item Memmott, J. et al. 1994. The structure of a tropical host-parasitoid community. J. An.Ecol. 63: 521-540.

\item Neck, R. W. 1976. Lepidopteran foodplant records from Texas. J. Res. Lep. 15(2): 75-82.

\item Nielsen, A. and Bascompte, J. 2007. Ecological networks, nestedness and sampling effort. J. Ecol. 95: 1134-1141.

\item Pielou, E. C. 1974. Biogeographic range comparisons and evidence of geographic variation in host-parasite relations. Ecology 55: 1359-1367.

\item Pipkin, S. B. et al. 1966. Plant host specificity among flower-feeding neotropical drosophila (Diptera: Drosophilidae). Am. Nat. 100: 135-156.

\item Podluss\'any, A. et al. 2001. On the leguminous host plants of seed predator weevils (Coleoptera: Apionidae, Curculionidae) in Hungary.  Acta Zoologica Academiae Scientiarum Hungaricae 47 (4): 285-299.

\item Prado, P. I. and Lewinsohn, T. M. 1994. Genus Tomoplagia (Diptera, Tephritidae) in the Serra do Cipó, MG, Brazil: Host Records and Notes of Taxonomic Interest. Revista Brasileira de Entomologia 38: 3-4.

\item Prado, P. I. et al. 2006. The nested structure of marine cleaning symbiosis: is it like flowers and
bees? Biology Letters 3 51-54.

\item Ratchke, B. J. 1976. Competition and coexistence within a guild of herbivorous insects. Ecology 57: 76-87.

\item Rodr\'{i}guez-Giron\'es, M. A. and Santamar\'{i}a, L. 2006. A new algorithm to calculate the nestedness temperature of presence-absence matrices. J. Biogeogr. 33: 924-935.

\item Selva, N. and Fortuna, M. 2007. The nested structure of a scavenger community. Proc. R. Soc. 274: 1101-1108.

\item Sheldon, J. K. and Rogers, L. E. 1978. Grasshopper food habitats within a shrub-steppe community. Oecologia 32: 85-92.

\item V\'azquez, D. P. et al. 2005. Species abundance and the distribution of specialization in host-parasite interaction networks. J. An. Ecol. 74: 946-955.

}

\end{description}

\break


\begin{center} {\Large Legend table}\end{center}

\begin{table}[h]
\label{tab1} \caption{Set of $23$ insect-plant community networks.
For each network we show : the occupancy $\rho$, size L$_1$ and
$\mbox L_2$, the nestedness estimator $\eta$ (multiplied by 100),
the temperature T and the reference in the literature.}
\end{table}

\break

\begin{center} {\Large Figure captions}\end{center}

\vspace{-.2cm}

\begin{figure}[!h]
\caption{Lattice representation of a particular lattice with
L$_1=\mbox{L}_2=5$ and N$=8$ (empty circles). The filled diamonds
show the maximal nested lattice for the same L$_1$, L$_2$ and N. We
remark that the circles represent the lattice after the packing
process, both P$_{\mbox {\footnotesize{L}}_1}$ and P$_{\mbox
{\footnotesize{L}}_2}$ are ranked.} \label{fig1}
\end{figure}

\vspace{-.2cm}

\begin{figure}[!h]
\caption{The behavior of d$_{\mbox{\scriptsize {min}}}$ and
d$_{\mbox{\scriptsize {rand}}}$ versus  $\rho$ for L$_1=$ L$_2=20$.
The curves obey the rule d$_{\mbox{\scriptsize {min}}} < $
d$_{\mbox{\scriptsize {rand}}}$ except for $\rho \rightarrow 1$.}
\label{fig2}
\end{figure}

\vspace{-.2cm}

\begin{figure}[!h]
\caption{The behavior of d$_{\mbox{\scriptsize {min}}}$ and
d$_{\mbox{\scriptsize {rand}}}$ versus L for a constant occupation
$\rho=0.5$. For both curves we have d $\propto$ L$^2$.} \label{fig3}
\end{figure}

\vspace{-.2cm}

\begin{figure}[!h]
\label{fig4} \caption{Four examples of real community matrices
referred to data files on table 1. In (a) the matrix corresponds to
file 10, in (b) we use file 9, in (c) file 2 and in (d) file 12.}
\end{figure}

\vspace{-.2cm}

\begin{figure}[!h]
\label{fig5} \caption{ The curve of occupancy $\rho$, versus
nestedness index $\eta$, for all the network tested in this work. }
\end{figure}

\break

\begin {flushleft}
\begin{tabular}{|c|c|c|c|c|l|} \hline
    \footnotesize File    &\footnotesize  Fill(\%) &\footnotesize Size (Plant $\times$ Insect)  &\footnotesize \footnotesize $T$    &\footnotesize $\eta \times 100$ & \footnotesize Reference \\
    \hline
   \footnotesize 1 &\footnotesize 42.9 &\footnotesize  18 $\times$ 57 &\footnotesize 28.0 &\footnotesize 17.2 &\footnotesize Futuyama and Gould 1979\\\hline
   \footnotesize 2 &\footnotesize 10.5 & \footnotesize 10 $\times$ 18 &\footnotesize 43.2 &\footnotesize 79.1 &\footnotesize Dawah et al. 1995 \\\hline
   \footnotesize 3 &\footnotesize 20.0 &\footnotesize 43 $\times$ 14 &\footnotesize 10.0 &\footnotesize 24.3 &\footnotesize Buruga and Olembo 1971 \\\hline
   \footnotesize 4 & \footnotesize8.4 &\footnotesize 63 $\times$ 25 & \footnotesize6.9 &\footnotesize 28.1 &\footnotesize Buruga and Olembo 1971 \\\hline
   \footnotesize 5 & \footnotesize8.6 &\footnotesize 33 $\times$ 55 &\footnotesize 13.5 &\footnotesize 40.8 &\footnotesize Claridge and Wilason 1981 \\\hline
    \footnotesize6 & \footnotesize1.1 &\footnotesize 107 $\times$ 104 &\footnotesize 7.29 &\footnotesize 83.5 &\footnotesize Flowers and Janzen 1997 \\\hline
    \footnotesize7 & \footnotesize16.0 &\footnotesize 52 $\times$ 22 &\footnotesize 20.7 &\footnotesize 35.1 &\footnotesize Joern 1979 \\\hline
    \footnotesize8 & \footnotesize13.3 &\footnotesize 54 $\times$ 24 & \footnotesize11.0 &\footnotesize 32.3 &\footnotesize Joern 1979 \\\hline
    \footnotesize9 &\footnotesize 34.7 &\footnotesize 11 $\times$ 11 &\footnotesize 9.4 &\footnotesize 23.0 &\footnotesize Pielou 1974 \\\hline
    \footnotesize10 &\footnotesize 43.5 &\footnotesize 13 $\times$ 12 &\footnotesize 19.3 &\footnotesize 18.1 &\footnotesize Pielou 1974 \\\hline
    \footnotesize11 & \footnotesize4.7 &\footnotesize 52 $\times$ 27 & \footnotesize14.5 &\footnotesize 62.7 &\footnotesize Jermy and Szentesi 2003 \\ \hline
    \footnotesize12 & \footnotesize2.0 &\footnotesize 53 $\times$ 92 & \footnotesize8.3 &\footnotesize 75.9 &\footnotesize Memmot et al. 1994 \\\hline
    \footnotesize13 &\footnotesize 5.5 &\footnotesize 46 $\times$ 22 &\footnotesize 20.1 &\footnotesize 62.8 &\footnotesize Neck 1976 \\\hline
    \footnotesize14 &\footnotesize 14.7 &\footnotesize 08 $\times$ 11 &\footnotesize 42.1 & \footnotesize68.1 &\footnotesize Pipkin et al. 1966 \\\hline
    \footnotesize15 &\footnotesize 4.4 &\footnotesize 55 $\times$ 43 &\footnotesize 8.3 & \footnotesize53.5 &\footnotesize Podluss\'any et al. 2001 \\\hline
    \footnotesize16 &\footnotesize 23.0 &\footnotesize 13 $\times$ 09 &\footnotesize 35.8 & \footnotesize42.3 &\footnotesize Ratchke 1976 \\\hline
    \footnotesize17 &\footnotesize 52.3 &\footnotesize 15 $\times$ 08 &\footnotesize 17.2 & \footnotesize18.2 &\footnotesize Sheldon and Rogers 1978 \\\hline
    \footnotesize18 & \footnotesize15.9 &\footnotesize 18 $\times$ 15 & \footnotesize26.3 &\footnotesize 45.3 &\footnotesize Prado and Lewinsohn 1994 \\\hline
    \footnotesize19 &\footnotesize 4.5 &\footnotesize 33 $\times$ 29 & \footnotesize20.5 & \footnotesize69.7 & \footnotesize Valladares, Argentina (unpublished) \\\hline
    \footnotesize20 &\footnotesize 8.9 &\footnotesize 27 $\times$ 22 & \footnotesize17.8 &\footnotesize 53.7 & \footnotesize Lewinsohn, Brazil (unpublished) \\\hline
    \footnotesize21 &\footnotesize 8.0 &\footnotesize 30 $\times$ 34 & \footnotesize14.1 & \footnotesize44.1 & \footnotesize Lewinsohn, Brazil (unpublished) \\\hline
    \footnotesize22 &\footnotesize 12.8 &\footnotesize 33 $\times$ 21 & \footnotesize17.7 &\footnotesize 37.9 & \footnotesize Almeida, Brazil (unpublished) \\\hline
    \footnotesize23 &\footnotesize 19.6 &\footnotesize 21 $\times$ 32 &\footnotesize 20.6 &\footnotesize 29.3 &\footnotesize Kubota, Brazil (unpublished)    \\\hline

 \end{tabular}

\end {flushleft}
\vspace{0.7 cm}
\begin{center}
{\Large Table 1}
\end {center}

 \break

\begin{figure}[ht]
\begin{center}
 \includegraphics[width=150mm]{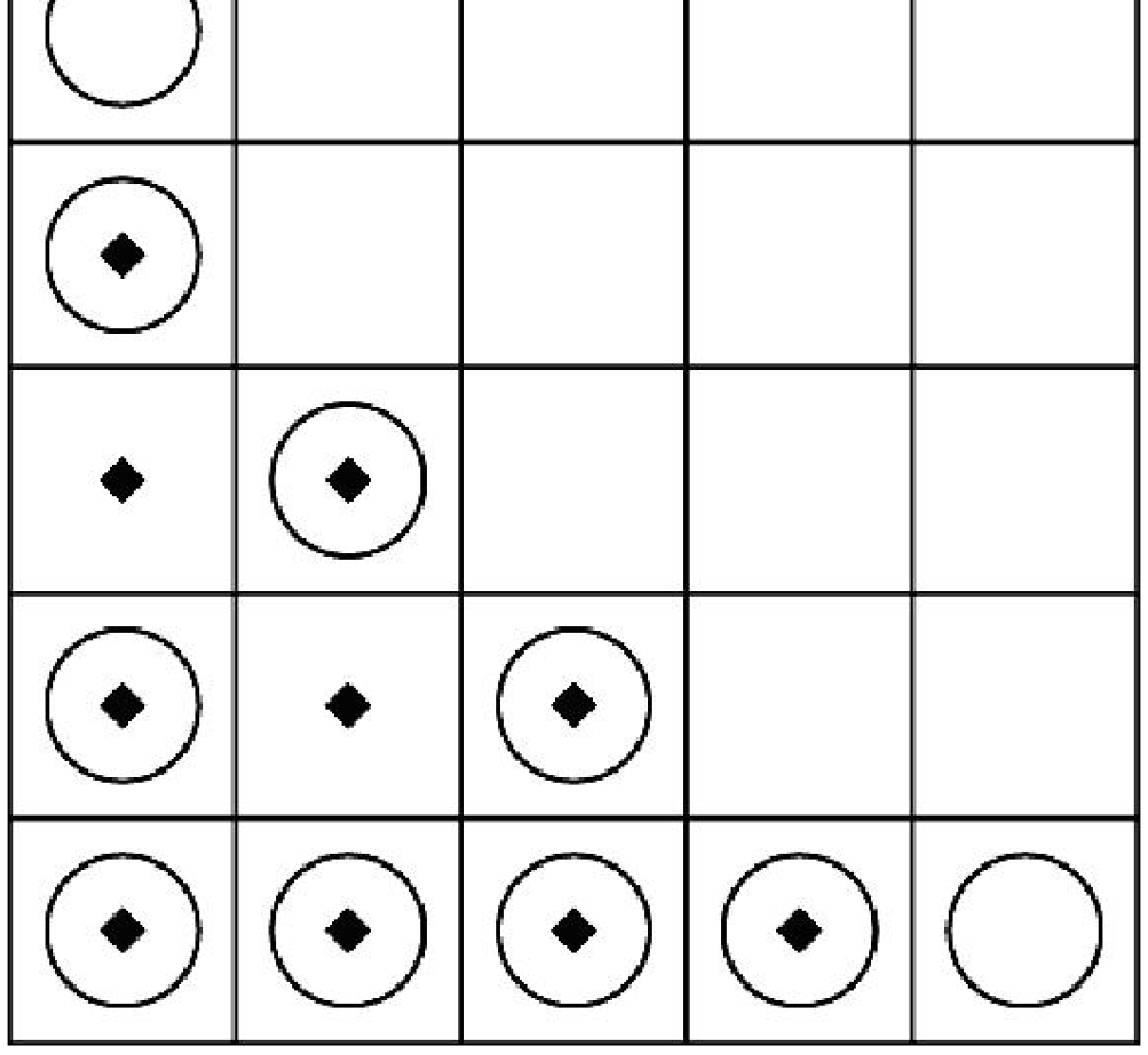}
\vspace{0.7cm}

{\Large Figure 1}
\end{center}
\vspace{2mm}
\end{figure}

\newpage

\begin{figure}[!h]
\begin{center}

\includegraphics[width=150mm]{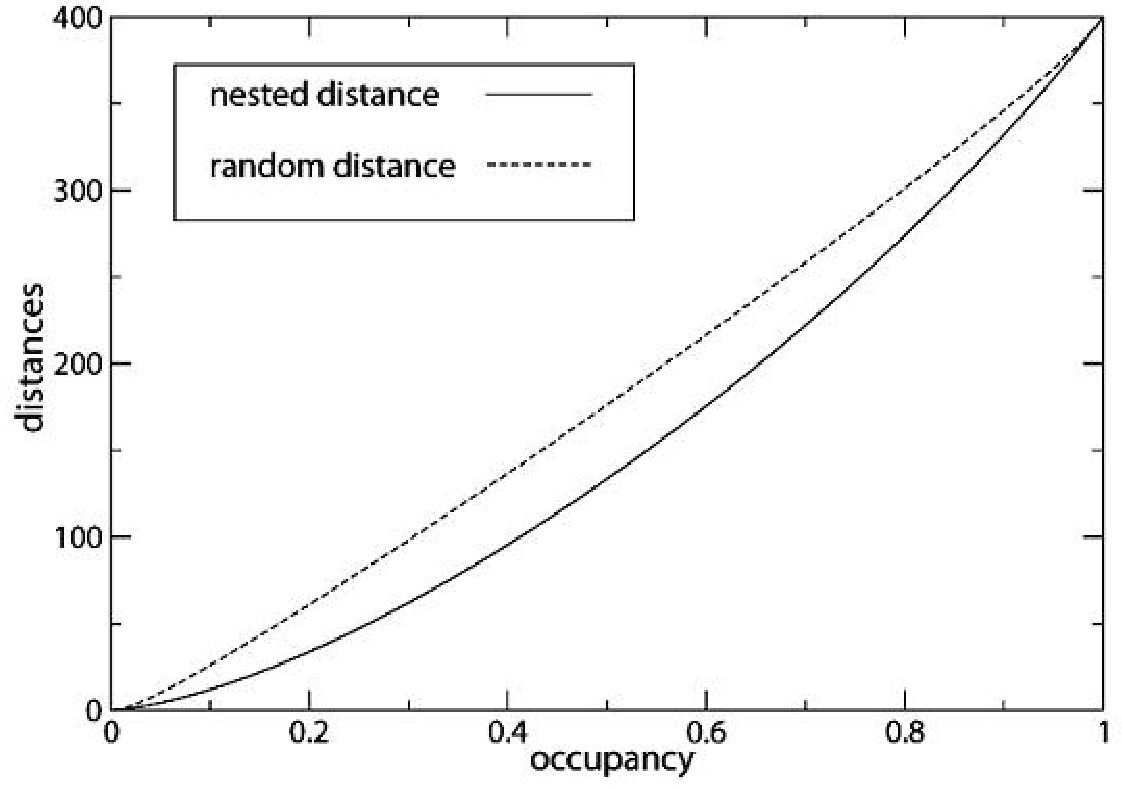}

{\Large Figure 2}
\end{center}
\vspace{2mm}
\end{figure}
\newpage

\begin{figure}[!h]
\begin{center}
 \includegraphics[width=170mm]{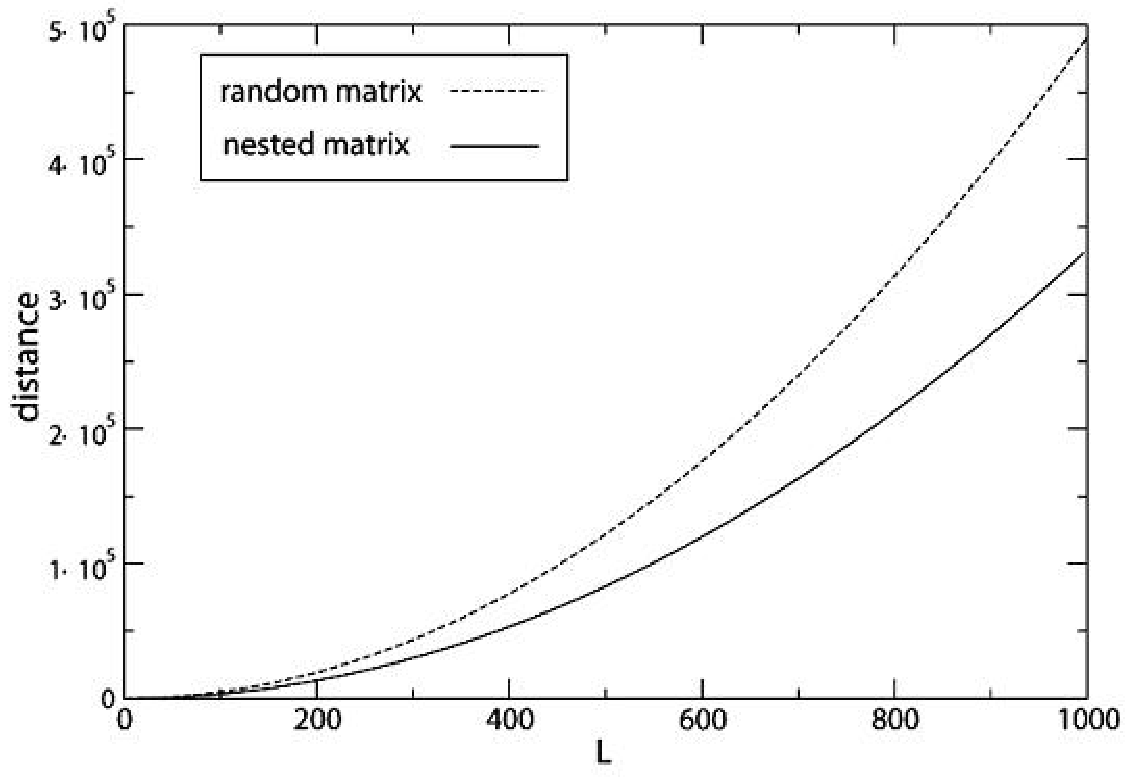}
{\Large Figure 3}
\end{center}
\vspace{2mm}
\end{figure}
\newpage

\begin{figure}[h]
\begin{center}
\resizebox{120mm}{!}{ \includegraphics[scale=0.2,width=80mm]{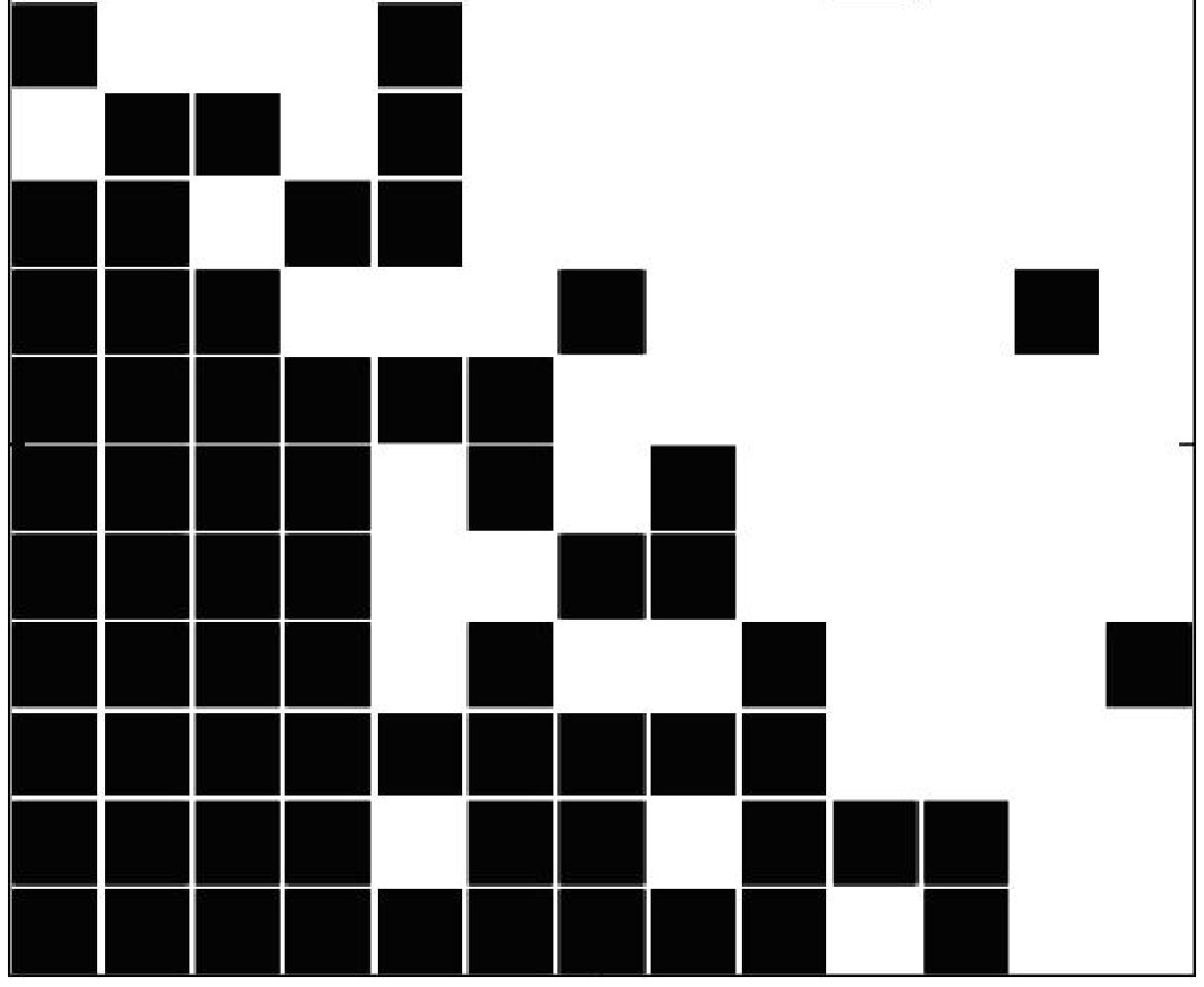}}
\vspace{0.7cm}

 {\Large Figure 4a}
\end{center}
\vspace{2mm}
\end{figure}

\newpage

\begin{figure}[h]
\begin{center}
\resizebox{120mm}{!}{ \includegraphics[angle=0]{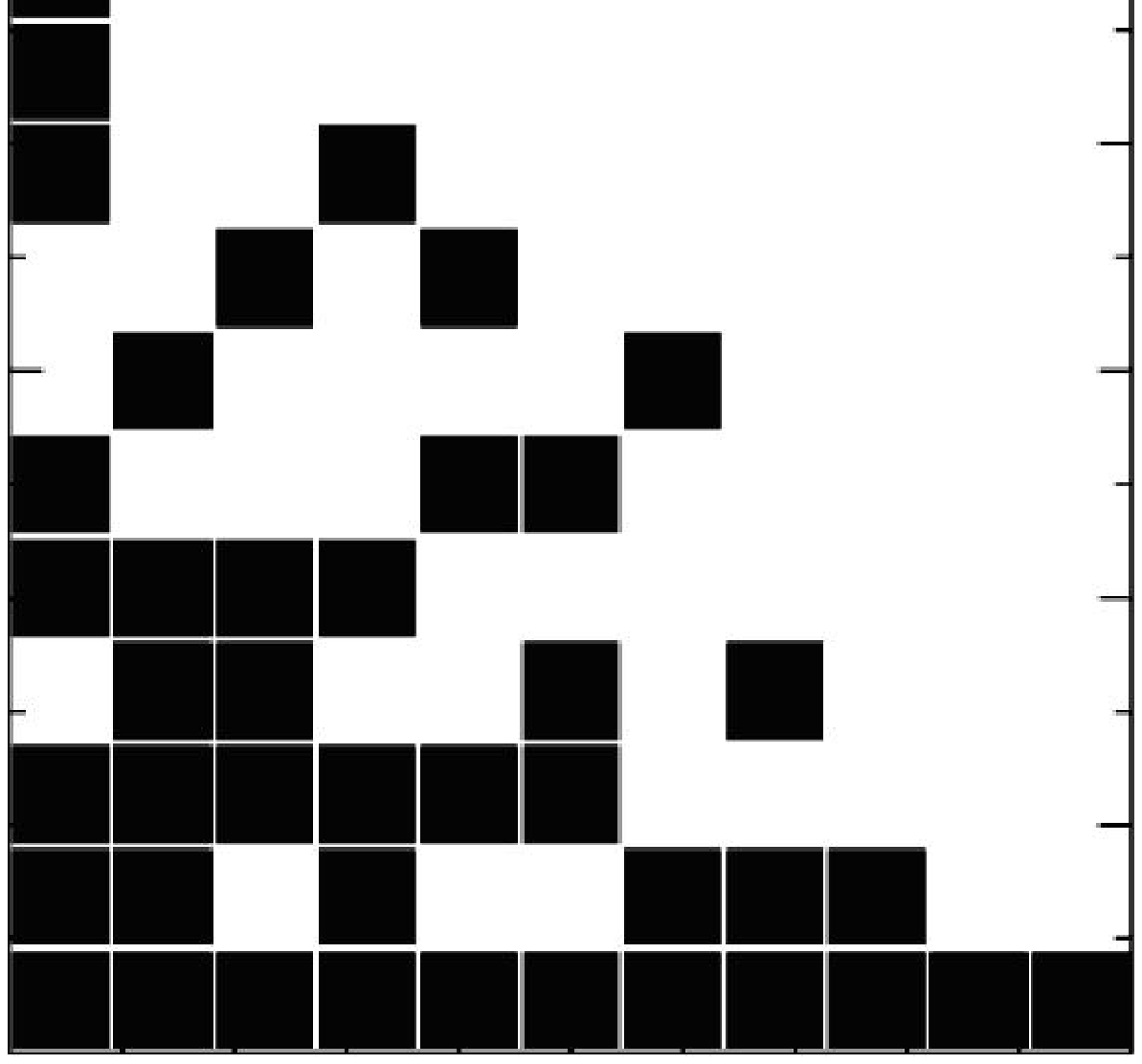}}
\vspace{0.7cm}

 {\Large Figure 4b}
\end{center}
\vspace{2mm}
\end{figure}

\newpage

\begin{figure}[h]
\begin{center}
\resizebox{120mm}{!}{ \includegraphics[angle=0]{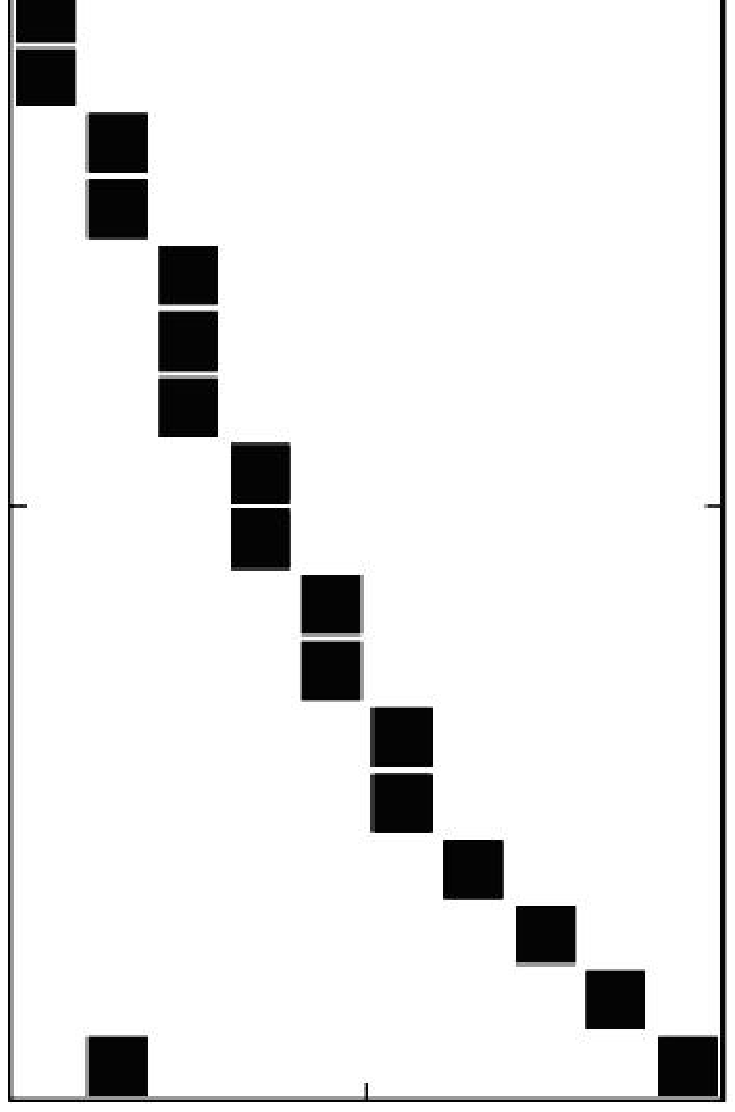}}
\vspace{0.7cm}

 {\Large Figure 4c}
\end{center}
\vspace{2mm}
\end{figure}

\newpage

\begin{figure}[h]
\begin{center}
\resizebox{120mm}{!}{ \includegraphics[angle=0]{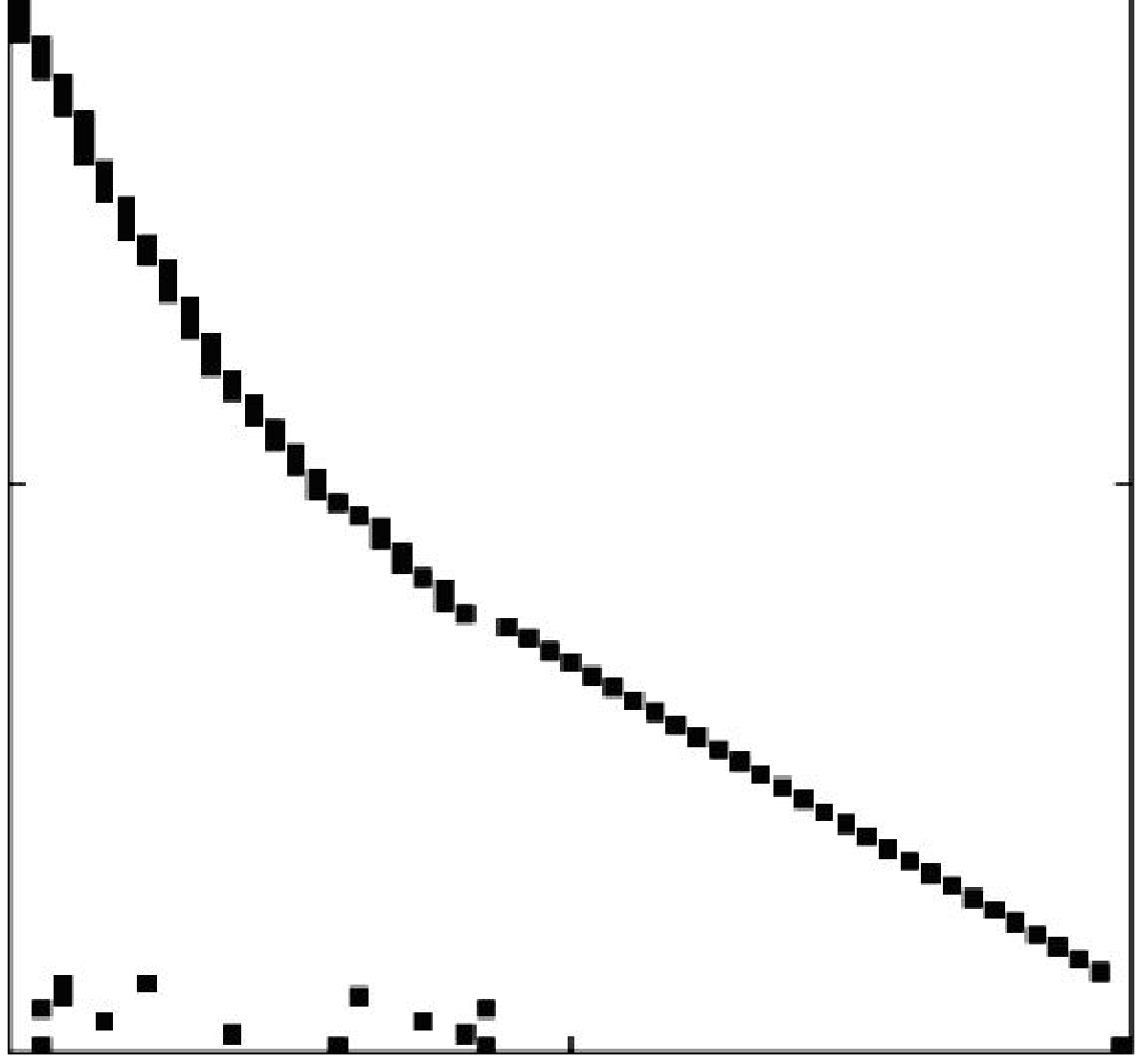}}
\vspace{0.7cm}

 {\Large Figure 4d}
\end{center}
\vspace{2mm}
\end{figure}

\newpage

\break

\begin{figure}[h]
\begin{center}
\resizebox{170mm}{!}{ \includegraphics[angle=0]{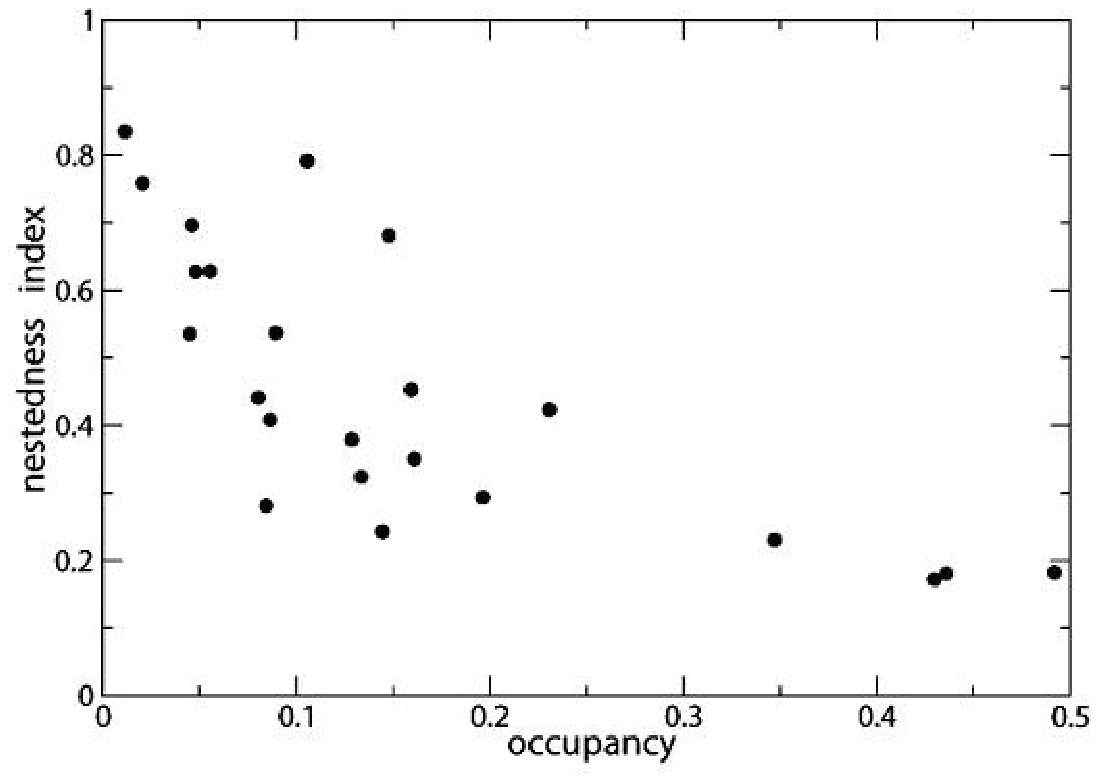}}

{\Large Figure 5}
\end{center}
\vspace{2mm} \label{figure 5}
\end{figure}

\end{document}